\documentclass[11pt]{article}
\usepackage[utf8]{inputenc}
\usepackage{amsmath,amssymb,amsthm}
\usepackage{bm,bbm,xfrac,url,hyperref,authblk}
\usepackage{fullpage}
\usepackage{booktabs}

\usepackage{subcaption}

\usepackage{todonotes}

\newcommand{\set}[1]{\left\lbrace #1 \right\rbrace}
\newcommand{\card}[1]{\left\lvert{#1}\right\rvert}
\newcommand{\absv}[1]{\card{#1}}

\newcommand{\contributor}{n}
\newcommand{\Contributor}{N}
\newcommand{\data}[1]{\mathcal D_{#1}}
\newcommand{\rating}{r}
\newcommand{\Video}{V}
\newcommand{\video}{v}
\newcommand{\videobis}{w}
\newcommand{\localscore}[1]{\theta_{#1}}
\newcommand{\globalscore}[1]{\rho_{#1}}
\newcommand{\datafamily}{\vec{\data{}}}
\newcommand{\localscorefamily}{\vec{\localscore{}}}
\newcommand{\lossperinput}{\ell}

\newcommand{\weight}[1]{w_{#1}}
\newcommand{\collabweight}{\lambda}
\newcommand{\globalweight}{\nu}
\newcommand{\Loss}{\textsc{Loss}}
\newcommand{\weightConstant}{C}
\newcommand{\Rating}{R}

\begin{document}

\title{Tournesol: A quest for a large, secure and trustworthy database \\ of reliable human judgments}

\author[1,2]{Lê-Nguyên Hoang}
\author[2]{Louis Faucon}
\author[2]{Aidan Jungo}
\author[2]{Sergei Volodin}
\author[1,2]{Dalia Papuc}  
\author[1,2]{Orfeas Liossatos} 
\author[3]{Ben Crulis} 
\author[2,4]{Mariame Tighanimine}
\author[2]{Isabela Constantin}
\author[1,2]{Anastasiia Kucherenko}
\author[2,5]{Alexandre Maurer}
\author[1,2]{Felix Grimberg}
\author[2,6]{Vlad Nitu}
\author[2]{Chris Vossen} 
\author[1,2]{Sébastien Rouault}
\author[2,7]{El-Mahdi El-Mhamdi}

\affil[1]{IC, EPFL, Switzerland}
\affil[2]{Tournesol Association, Switzerland}
\affil[3]{University of Tours, France}
\affil[4]{LISE, CNAM-CNRS, France}
\affil[5]{UM6P, Benguerir, Morocco}
\affil[6]{CNRS, INSA Lyon, France}
\affil[7]{{\'E}cole Polytechnique, France}

\date{}
\maketitle

\begin{abstract}
Today's large-scale algorithms have become immensely influential, as they recommend and moderate the content that billions of humans are exposed to on a daily basis. These algorithms are the de-facto regulators  of the information diet of billions of humans, from shaping opinions on public health information to organizing groups for social movements.
This creates serious concerns, but also great opportunities to promote quality information~\cite{Hoang2020,HoangFEM2021}.
Addressing the concerns and seizing the opportunities is a challenging, enormous and fabulous endeavor~\cite{HoangElMhamdi19}, as intuitively appealing ideas often come with unforeseen unwanted {\it side effects}~\cite{ElMhamdiHoang21}, and as it requires us to think about what we truly and deeply prefer~\cite{soares2015value}.

To make progress, it is critical to understand how today's large-scale algorithms are built, and to determine what interventions will be most effective.
Given that these algorithms rely heavily on {\it machine learning}, we make the following key observation: \emph{any algorithm trained on uncontrolled data must not be trusted}.
Indeed, a malicious entity could take control over the data, poison it with dangerously misleading or manipulative fabricated inputs, and thereby make the trained algorithm extremely unsafe.
We thus argue that the first step towards safe and ethical large-scale algorithms must be the collection of a large, secure and trustworthy dataset of reliable human judgments.

To achieve this, we introduce \emph{Tournesol}, an open source platform available at \url{https://tournesol.app}.
Tournesol aims to collect a large database of human judgments on what algorithms ought to widely recommend (and what algorithms ought to stop widely recommending). 
In this paper, we outline the structure of the Tournesol database, the key features of the Tournesol platform and the main hurdles that must be overcome to make it a successful project.
Most importantly, we argue that, if successful, Tournesol may then serve as the essential foundation for any safe and ethical large-scale algorithm.
\end{abstract}

\newpage
\section{Introduction}

Algorithms are becoming increasingly influential. 
This is particularly the case of large-scale content recommendation algorithms, conversational algorithms and moderation algorithms.
For example, in 2019, 70\% of YouTube's one-billion hours of daily views resulted from algorithmic recommendations~\cite{cnet18}.
Meanwhile, in China, Microsoft's conversational algorithm called Xiaoice is used by over 600 million humans~\cite{ZhouJDH20,Durden20}. 
Finally, Facebook had to remove 6 billion fake accounts from their platforms within just one year~\cite{facebook_fake_accounts},  
mostly through algorithmic means.
In this context, \cite{HoangElMhamdi19} argues that, unless massive financial and cognitive investments are made to ensure that such algorithms are robustly beneficial, they will inevitably have unwanted large-scale side effects through unsafe information dissemination.
On the other hand, enormous opportunities in terms of quality information~\cite{Hoang2020} and public health~\cite{HoangFEM2021} could be seized if algorithms were designed with careful consideration for their societal impact ~\cite{VendrovNixon19,StrayAH20}.

Current leading concerns of unethical conversational, moderation and recommendation algorithms include cyberbullying~\cite{HindujaPatchin10}, fairness~\cite{MehrabiMSLG19}, discrimination~\cite{AbidFZ21}, privacy~\cite{TanuwidjajaCK19,CarliniTWJH+20}, job displacement~\cite{MartensTolan18}, radicalization \cite{RibeiroOWAM20,McGuffieNewhouse20}, political manipulation \cite{WoolleyHoward+18,woolley20}, misinformation \cite{WuMCL19,Aral20}, mute news (information that is drowned within the flood of information) \cite{RoslingRonnlundRR19}, information overload~\cite{Roetzel19}, anger \cite{BergerMilkman12}, hate \cite{SchmidtWiegand17}, geopolitical tensions~\cite{AlexanderKS21}, addiction~\cite{TurelBB18,HawiSamaha16}, inability to focus \cite{MarkICJS16}, mental health \cite{EichstaedtSMUC+18}, autonomous weapons \cite{RussellHAV15}, arms race \cite{Geist16} and existential risk \cite{Bostrom13}.

\paragraph{The intermediate body crisis.}

As societies grew in size and complexity~\cite{flannery1972cultural}, {\it intermediate bodies} have taken an increasingly central role, to mediate the interactions between individuals.
Such intermediate bodies include lawmakers and states~\cite{Durkheim2014division}, but also journalists, teachers and public health officials, among many others. 
At the heart of the role of each of these groups is information processing in order to judge what messages ought to be moderated, communicated or amplified.

As online activities grew, today's digital platforms have de-facto taken the role that was traditionally played by these intermediate bodies~\cite{tighanimine2019, hadfield2017rules}. 
This became particularly striking when, in 2020, the then President of the United States was banned from Twitter, Facebook, and Youtube, long before any court sentenced him for inciting violence in Capitol Hill. 
As another example, during the COVID-19 pandemic, digital platforms had to make decisions on health misinformation, often months before traditional intermediate bodies such as WHO, the CDC, and others.
Arguably, traditional intermediate bodies currently lack the tools to play the important online role they had offline.

Tournesol's ambition is to empower intermediate bodies by building algorithms that make \emph{scalable} judgments\footnote{We recall, as in~\cite{HoangElMhamdi19} that the very word ``Algorithm'' was coined after a lawyer, Alkhawarizmi, whose ambition in inventing Algebra was to help other lawyers make better judgments in courts by making decision making operations such as inheritance cases' transparent, repeatable and corrigible, through step-by-step guidelines. In the light of this remark, Algorithmic decision making should be viewed as an asset for society, rather than an obscure tool.} which better represent what they would recommend to society.
We hope to enable them to effectively mediate online human interactions, at the rate of billions of decisions per minute, as required by social medias.

\paragraph{Data is key.} Importantly, today's large-scale algorithms heavily rely on \emph{machine learning}.
In other words, such algorithms leverage massive amounts of data to achieve high performance~\cite{FedusZS21}.
However, an immediate corollary of this is that such algorithms are \emph{manipulated} by their data, especially given their current designs.
If the data is full of discriminatory texts, then such ``stochastic parrots'' will repeat or favor such discriminatory texts~\cite{BenderGMS21}.
What is especially concerning is that today's most influential algorithms are mostly powered by \emph{user-generated} data downloaded from the web~\cite{SmithSPKCL13,WangSMHLB19,WangPNSMHLB19}.
As a result, many of today's trained algorithms should be considered  extremely unsafe.
More generally, \emph{any algorithm trained on uncontrolled data must not be trusted}.

A recent line of research on Byzantine resilient learning algorithms provides new tools to protect algorithms from poisoned data or malicious actors~\cite{BlanchardMGS17,MhamdiGR18,MhamdiGR20, el2020robust, sebastien}. 
However, such protections cannot perform miracles.
Especially in heterogeneous environments, impossibility theorems apply~\cite{MhamdiFGGHR21}.
In particular, there can be no algorithmic trick to protect against a majority of maliciously crafted data sources.

More generally, the safety of learning algorithms demands the safety of their training datasets.
In particular, to design trustworthy ethical algorithms, it is essential to train them on large, secured and trustworthy datasets of human ethical judgments.
To the best of our knowledge, today, there is no such dataset.
As a result, according to our reasoning, there can currently be no safe learning algorithm.
The goal of Tournesol, available at \url{https://tournesol.app}, is to remedy this state of affairs, by constructing the largest, most secured and most trustworthy dataset of reliable human judgments ever created.

\begin{figure}[!ht]
    \centering
    \includegraphics[width=.9\linewidth]{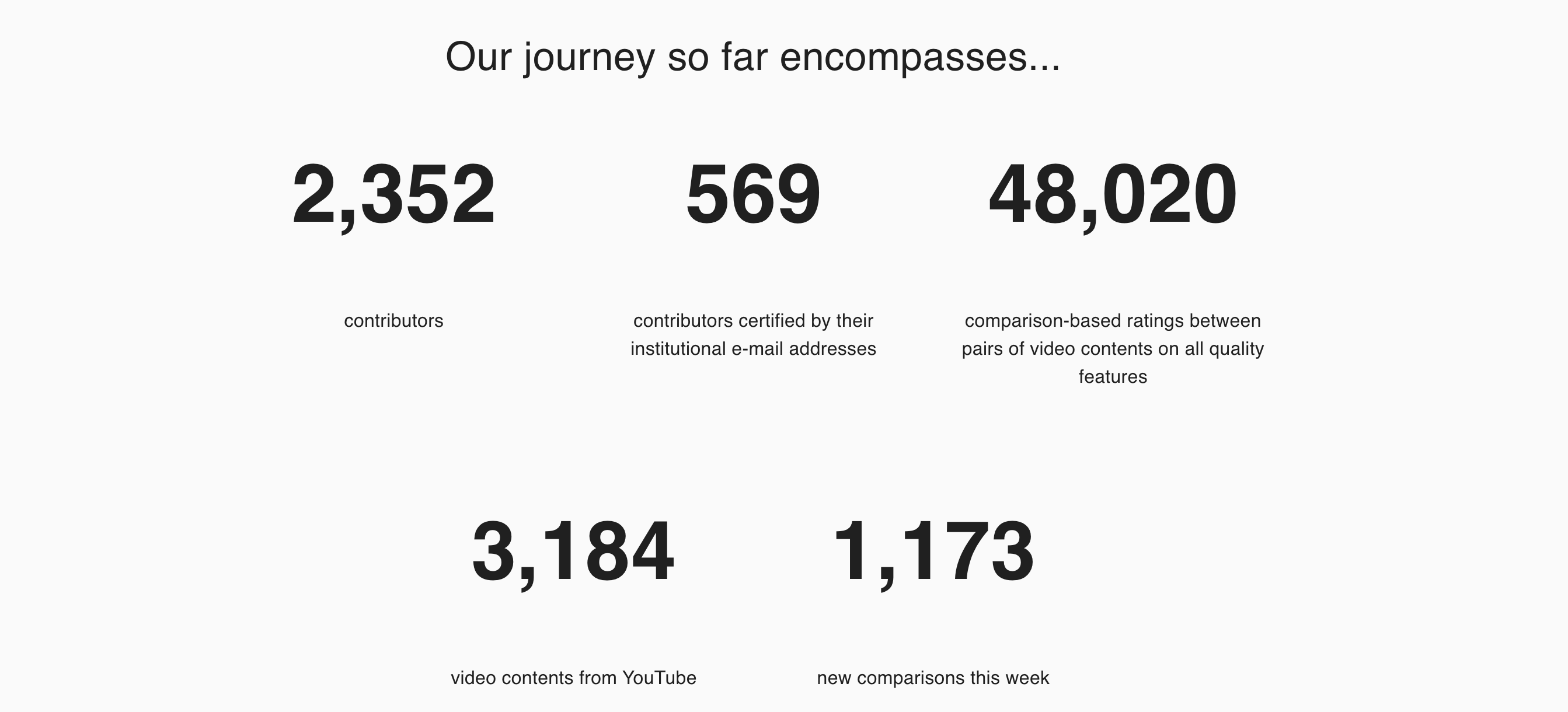}
    \caption{Global statistics of Tournesol on May 27, 2021}
    \label{fig:global_statistics}
\end{figure}

\paragraph{Introducing Tournesol.} More specifically, Tournesol aims to elicit comparison-based judgments on what videos are preferable to recommend widely, and on quality features such as ``reliable and not misleading'', ``important and actionable'' and ``engaging and thought-provoking'', among others. 

These judgments are then leveraged by a machine learning model, which infers a global score for each video. The model roughly combines the Bradley-Terry model~\cite{BradleyTerry52,Maystre18} and the Licchavi framework~\cite{FarhadkhaniGH21}, which allows to provide Byzantine resilience and (approximate) strategyproofness. Byzantine resilience means that the outputs of the model are resilient to a minority of contributors with arbitrary behaviors, while strategyproofness means that it is in strategic contributors' best interests to provide judgments that match their true preferences. We stress that strategyproofness is particularly important to guarantee as much as possible the trustworthiness of our dataset. 

Moreover, contributors are asked to provide personal information, which we use to assess their trustworthiness. In particular, we rely on email verification from trusted email domains, to protect Tournesol against Sybil attacks, i.e., attackers creating and having multiple identities~\cite{Douceur02}. In the near future, we also plan to deploy a vouching mechanism to include more contributors without sacrificing too much security. Contributors are also asked to report their degrees and expertise, as well as demographic data.

Tournesol's interface allows contributors to either provide data publicly or privately. In the former case, their data will be appended to the Tournesol public database, which is readily downloadable from the Tournesol home page\footnote{\url{https://tournesol.app/tournesol.public.latest.csv.zip}}. We encourage its widespread reuse and remix, given appropriate attribution,  under the conditions of the license CC BY-SA\footnote{\url{https://creativecommons.org/licenses/by-sa/2.0/}}. 
Moreover, we ask that the entities who reuse the data pay great attention to strategyproofness considerations, for the trustworthiness of the Tournesol database depends on this. 
Finally, we ask any entity that reuses our data to do so responsibly, and, as much as possible, for the good of all of humanity.
In particular, we ask them to avoid reusing our data for, e.g. advertisement targeting, especially if what is advertised is not for the good of our contributors or of humanity. 

Tournesol is deeply attached to transparency. 
Our code is open-source\footnote{\url{http://github.com/tournesol-app/}}, and nearly all of our discussions on bugs and new functionalities are publicly available\footnote{\url{https://discord.gg/3KApy6tHAM}}. 
Our front-end uses React~\cite{gackenheimer2015}, and the back-end relies on Django~\cite{holovaty2009}, with machine learning algorithms running with Tensorflow~\cite{abadi2016}. 
Our server's statistics are also public\footnote{\url{https://munin.tournesol.app/}}. 
Note also that the Tournesol platform is still rapidly evolving. 
The present document will be updated accordingly every few months. We refer however to the Tournesol wiki, available at \url{https://wiki.tournesol.app}, for more frequently updated information. 

\paragraph{Structure of the paper.} The rest of the paper is organized as follows. 
Section~\ref{sec:database} will present the key elements of our databases, and our motivations to collect the data we chose to collect.
It will also briefly discuss our privacy policy and options.
Section~\ref{sec:algorithm} then presents the algorithm currently in use on Tournesol. 
We stress in particular that our algorithm, based on Licchavi~\cite{FarhadkhaniGH21}, obeys the principle ``one contributor, one unit force'', which makes it somewhat strategyproof.
More importantly, we hope that this section will motivate and inspire new research into analyzing user-generated comparison-based judgment datasets like Tournesol's.
Section~\ref{sec:statistics} presents our preliminary analysis of our dataset.
Since the data collection process is ongoing, we stress that the statistics we present are bound to evolve within the next months and years.
Nevertheless, we hope to thereby give insights into both the data we have collected so far, and to illustrate the kind of insights that the analysis of our database can yield.
Section~\ref{sec:challenges} then provides a list of research challenges that we regard as urgent to tackle for safer and more ethical algorithms in general, and for the specific case of Tournesol in particular.
Finally, Section~\ref{sec:conclusion} provides a summary, as well as a call for contributions to make Tournesol a success, and to move towards safer and more ethical large-scale algorithms.

\section{The databases}
\label{sec:database}

In this section, we describe the main contribution of Tournesol, namely its new, scalable, secured and trustworthy database of reliable human judgments.

\subsection{Comparison-based judgments}
\label{sec:ratings}

To increase the quality of our data, we ask contributors to provide comparison-based judgments, by building upon a large literature on the value of such comparisons~\cite{Festinger54,BradleyTerry52,Maystre18}.

It is important to stress that the extent to which a piece of content should be recommended must not be measured by a binary variable.
Indeed, while some pieces of content are highly undesirable to recommend at all, others are somewhat desirable to recommend widely, and others still are very important to recommend at scale.
The same remark arguably holds for all the other quality criteria considered by Tournesol.
A video can ``encourage better habits'', be ``layman-friendly'', and promote ``diversity and inclusion'' to varying degrees.

This suggests that videos should be scored on a continuous scale for each quality criterion.
This is what Tournesol eventually proposes as an end product on its platform.
Contributors could have been asked to judge videos on a Likert scale~\cite{Likert32}.
But, while the Likert scale provides valuable information~\cite{Norman10,WillitsTL16}, the Likert scale has been widely criticized for being an unreliable measure and difficult to interpret~\cite{Albaum97,JoshiKCP15,Subedi16}.
In the case of Tournesol, we were concerned in particular that contributors abuse extreme values on the scale, or that different contributors use different values of the scale to mean different things~\cite{LeeJMZ02}.
Intuitively, for instance, if a math video is consistently reviewed by extremely rigorous contributors, it might end up with a worse ``reliability and not misleading'' score than a video on another topic that was reviewed by much less demanding contributors.
Moreover, we aim to obtain a much more fine-grained judgment of video quality.

\begin{figure}[!ht]
    \centering
    \includegraphics[width=\linewidth]{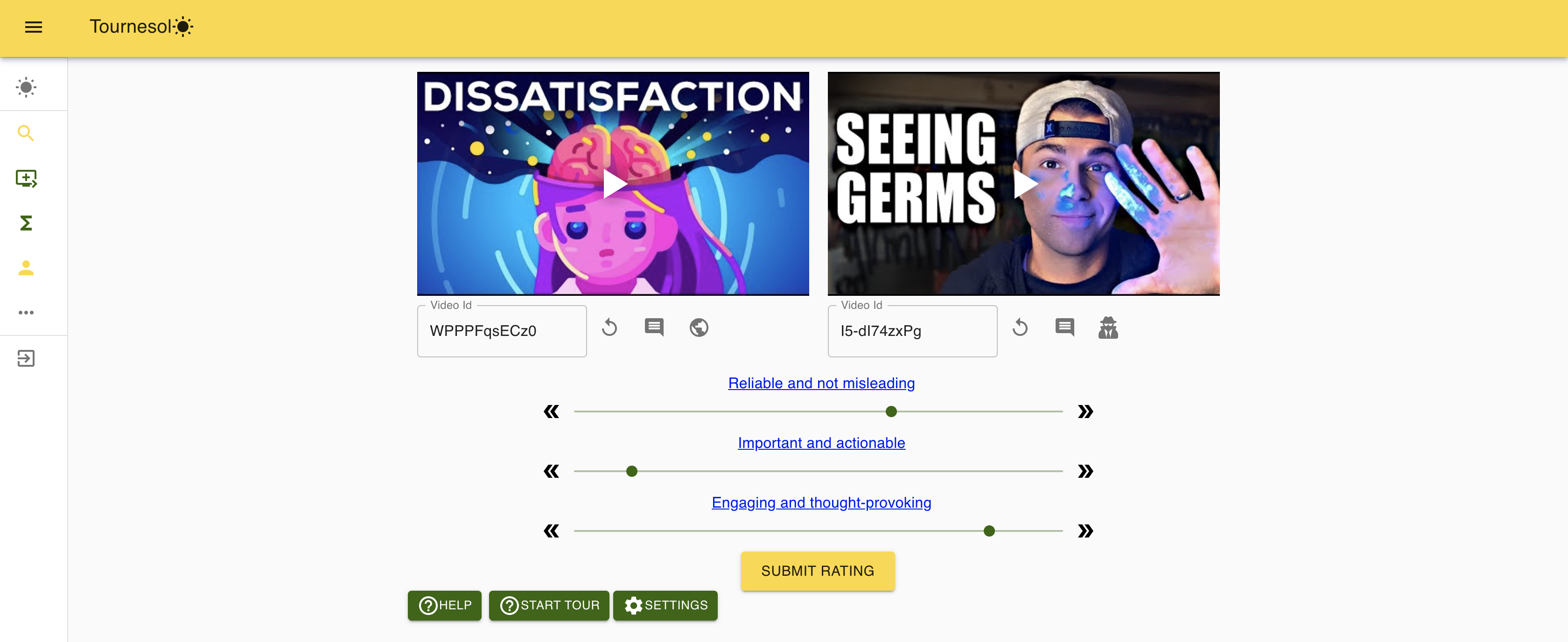}
    \caption{This is the Tournesol interface through which contributors are asked to provide judgments. The judgments are comparisons of different video content along different quality criteria.}
    \label{fig:rate-page}
\end{figure}

To circumvent the difficulty of extreme scoring and of incomparable individual scales, Tournesol asks contributors to provide comparison-based ratings (see Figure~\ref{fig:rate-page}).
Namely, contributors are asked to select two videos, and to tell Tournesol which one of the videos should be recommended at scale.
Moreover, rather than a binary decision, the contributor is asked to provide the judgment by moving a slider on a continuous scale, from $0$ to $100$.
When rating a quality criterion $Q$, the value $0$ means that the contributor judges that the left video content is far better in terms of $Q$, while the value $100$ means that they believe the right video content is far better in terms of $Q$.
Evidently, if the contributor reports $50$, it means that they put the slider in the middle, as they judge that the two videos should be scored similarly according to $Q$.

We believe that such comparative judgments will yield more reliable human judgments. 
We hope that this will allow for the design of more reliable algorithm training.

\subsection{Quality criteria}

Tournesol currently proposes ten \emph{quality criteria} to rate, only one of which is available by default.

The only \emph{default} quality criterion is ``Should be largely recommended''.
Tournesol chose to single out this one criterion as it is arguably the bottom line criterion for constructing recommendation algorithms.
Moreover, Tournesol chose to make it the only default quality criterion to facilitate the use of Tournesol for a large public of contributors.
We were particularly concerned about the risk of overwhelming new contributors with too many complex queries.

Nevertheless, Tournesol provides contributors with the possibility to rate nine other \emph{optional} quality criteria:
\begin{description}
\item[Reliable and not misleading:] Content that scores high on `reliable and not misleading' should make nearly all viewers improve their global world model, despite viewers' biases and motivated reasoning.
\item[Important and actionable:] Content that scores high on `important and actionable' should present data and arguments with major consequences, as well as actionable plans that would have a large impact.
\item[Engaging and thought-provoking:] Content that scores high on `engaging and thought-provoking' should catch the attention of a larger audience, trigger their curiosity and a desire to find out more or question their own beliefs.
\item[Encourages better habits:] Content that scores high on `encourages better habits' should be successful at motivating viewers to grow and improve themselves.
\item[Clear and pedagogical:] Content that scores high on `clear and pedagogical' should help viewers understand all the elements that lead to a conclusion.
\item[Layman-friendly:] Content that scores high on `layman-friendly' should be accessible to a very large audience.
\item[Diversity and inclusion:] Content that scores high on `diversity and inclusion' should celebrate diversity and be appealing to minority groups.
\item[Resilience to backfiring risks:] Content that scores high on `resilience to backfiring risks' should be safe to recommend to all sorts of viewers, with limited risks of misunderstandings.
\item[Entertaining and relaxing:] Content that scores high on `entertaining and relaxing' should entertain and relax viewers.
\end{description}
While detailed descriptions of all the criteria are provided on the Tournesol wiki, at \url{https://wiki.tournesol.app/index.php/Quality_criteria}, data scientists and researchers should probably not expect most contributors to read thoroughly our descriptions.
Arguably, most contributors will more likely judge these criteria according to their own understanding, which will be mostly based on the name of the criteria.

Note also that contributors can skip the judgment of a quality criterion.
The database is partially sparse in this regard.

\subsection{Confidence}

\begin{figure}[!ht]
    \centering
    \includegraphics[width=\linewidth]{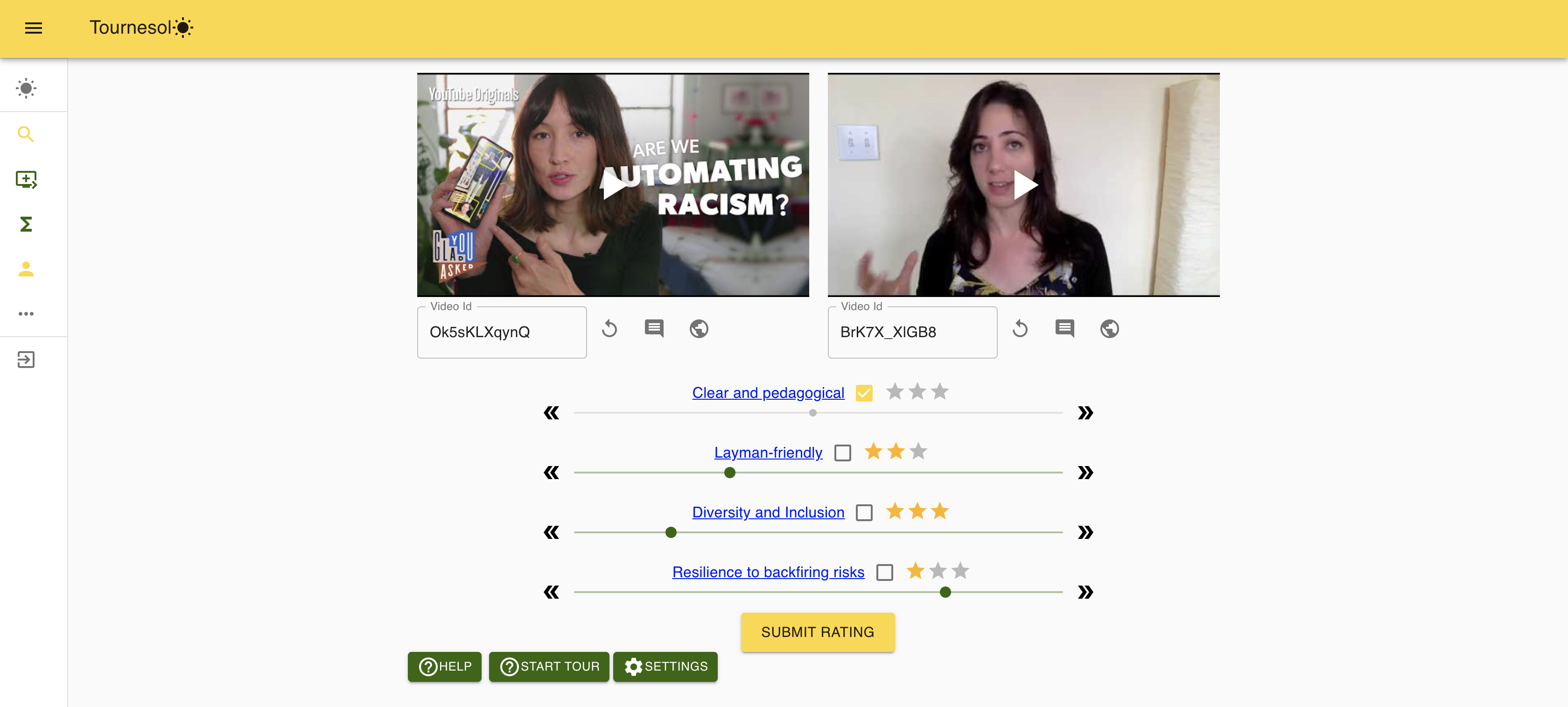}
    \caption{Tournesol allows contributors to judge videos, as well as the confidence they have in their judgments}
    \label{fig:confidence}
\end{figure}

Tournesol also proposes the option for contributors to assess their confidence in their ratings along each quality criterion, on a scale from $0$ to $3$, as illustrated in Figure~\ref{fig:confidence}.
A confidence of $0$ is equivalent to skipping the criterion altogether.

Such data are currently used by our learning algorithm to determine both the contributors' scores, and their impacts on the global Tournesol scores.

\subsection{Email verification}

To guarantee the security of our data, Tournesol aims to verify that every account is owned and controlled by a human, and that this human only owns and controls this single account on the platform.
In other words, Tournesol aims to obtain a \emph{Proof of Personhood}~\cite{BorgeKJGGF17} to verify each active Tournesol account, and to thereby prevent \emph{Sybil attacks}~\cite{Douceur02}.
Unfortunately, there is currently no reliable and scalable solution for \emph{Proof of Personhood}.

Today's main solution is \emph{email certification}.
More precisely, when they create a Tournesol account, contributors are asked to validate, if possible, an email address from a trusted email domain. 
The list of trusted email domains is currently managed manually.
An email domain will be considered trusted if it seems sufficiently unlikely that a large number of fake accounts can be created from this email domain.

Clearly, this excludes email provider domains like @gmail.com and personal domains like @my-personal-website.com.
Indeed, the concern is not only that the email domain owner will maliciously create a large number of fake accounts; it is also that they may be hacked by a malicious entity that will create such fake accounts.
The list of trusted email domains is available at \url{https://tournesol.app/email_domains}.
It includes domains like @epfl.ch, @who.int and @rsf.org.

Evidently, however, this solution is still highly imperfect.
On one hand, this does not guarantee the absence of fake accounts.
On the other hand, and perhaps more importantly, this excludes most potential contributors from participating.

\subsection{Vouching mechanism}
\label{sec:vouch}

\begin{figure}[!ht]
    \centering
    \includegraphics[width=.5\linewidth]{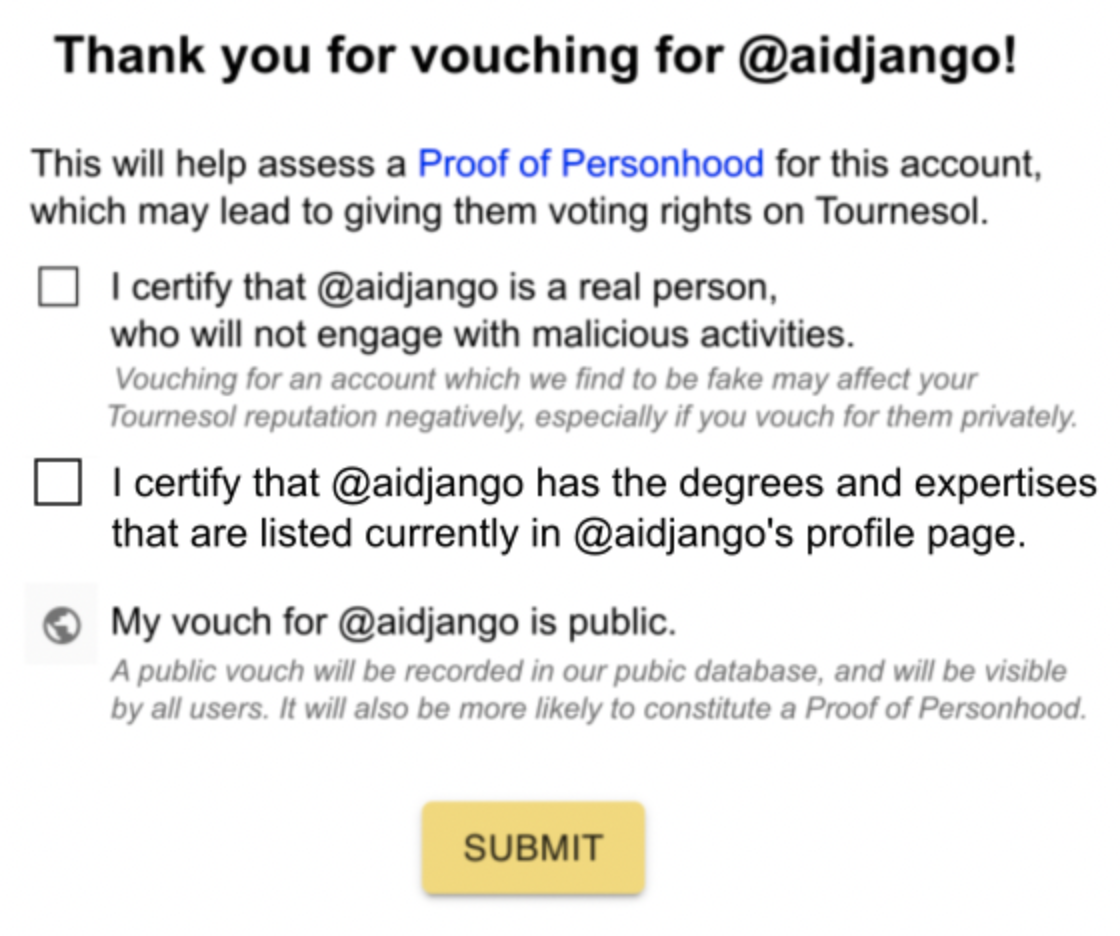}
    \caption{Any account on Tournesol can vouch for another account. This helps Tournesol verify more accounts in a secured manner.}
    \label{fig:vouch}
\end{figure}

Tournesol also proposes a vouching mechanism (see Figure~\ref{fig:vouch}). 
Namely, any account can vouch for the authenticity of another account. 
More precisely, the account must vouch that the other account is used by a human who is not using any other account on the platform.
Each email-verified account is then given a certain amount of vouching power, and each account must receive a certain amount of vouching (weighted by vouching power) to be certified.
We refer to \url{https://wiki.tournesol.app/index.php/Vouching_mechanism} for more (updated) details.

We are currently investigating the design of a reliable, Byzantine-resilient and scalable vouching mechanism (see Section~\ref{sec:pop}).

\subsection{Meta-data}
\label{sec:meta-data}

To better understand our contributors' rating patterns, and identify which ratings follow from thoughtful reflections, Tournesol measures the contributors' response times.
Tournesol also records the motions of the sliders.
We believe this to be particular important for volition learning (see Section~\ref{sec:volition}).

\subsection{Privacy}

Contributors can provide ratings publicly or privately.
More precisely, each contributor can select the privacy setting of any video they rate.
If a video is rated privately, then all its comparisons to any other video will be recorded privately; namely, only Tournesol's server will have access to such data.
Conversely, all comparisons that involve two publicly rated videos are public, and can be downloaded from the Tournesol home page.

Overall, we encourage transparency in our contributors, as we believe that this will foster important research on human judgments, and help make safer and more ethical algorithms.
However, we acknowledge that, because of social and political pressures, some judgments are dangerous to make public.
In such cases, Tournesol allows contributors to provide these judgments in a private way that nevertheless affects the Tournesol global scores, and thus what will be recommended by our platform.

\subsection{Personal information}

Each Tournesol contributor has a dedicated Tournesol contributor page, where they can access their individual scores for different videos, as well as statistics based on their recommendations.

On this page, we also ask contributors to provide personal information, publicly or privately.
Again, we encourage transparency from our contributors, as this will greatly facilitate research and the design of solutions for safer and more ethical algorithms.
But we acknowledge that this comes at a cost for our contributors, which they may want to avoid.

Personal information of great interest to Tournesol includes the contributor's expertise and degrees, as well as their demographic data.
Such data are particularly critical for Tournesol to understand its sampling bias, and to design debiasing solutions to avoid discriminatory algorithms.

\section{Tournesol's learning algorithm}
\label{sec:algorithm}

\begin{figure}[!ht]
    \centering
    \includegraphics[width=\linewidth]{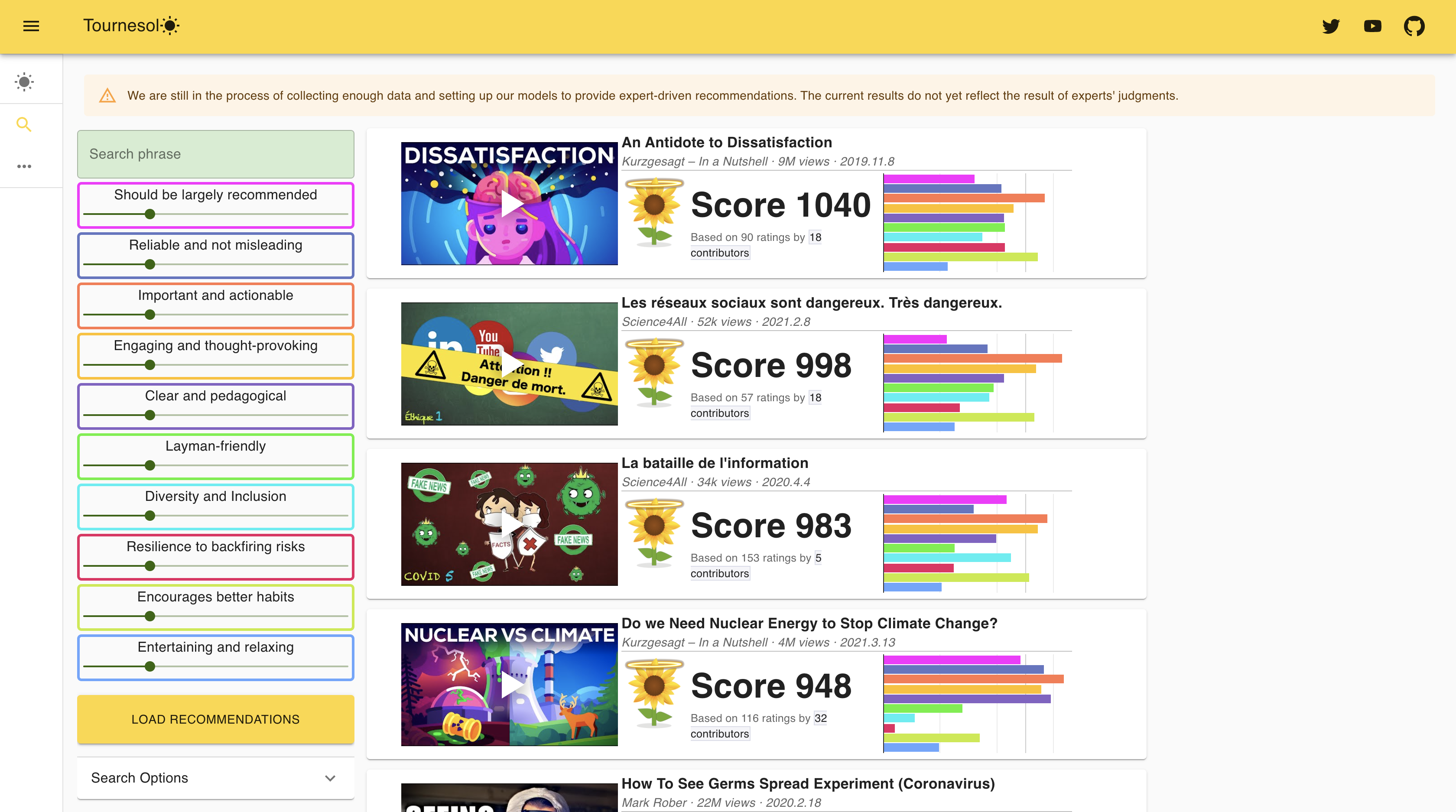}
    \caption{Top recommendations on May 27, 2021, on Tournesol's website. Note that, on the Tournesol website, users can customize their recommendations by adjusting the importance they assign to the different quality criteria.}
    \label{fig:top_recommendations}
\end{figure}

In this section, we briefly describe the learning algorithm used by Tournesol to transform comparison-based ratings into individual and global scores.
We hope that this will increase the transparency of Tournesol, inspire further research into leveraging our data and stress the importance we assign to strategyproof learning.
Figure~\ref{fig:top_recommendations} lists the current top recommendations on Tournesol.

Note that each quality criterion is treated independently.
Thus, without loss of generality, we only focus on a single criterion here.

\subsection{Licchavi}

Tournesol uses the Licchavi framework~\cite{FarhadkhaniGH21}.
More precisely, denote $[\Contributor] = \set{1, \ldots, \Contributor}$ the set of verified Tournesol contributors, and $[\Video] = \set{1, \ldots, \Video}$ the set of rated videos.
We denote $\data{\contributor} = \set{(\video, \videobis, \rating)}$ a set of ratings by contributor $\contributor \in \Contributor$.
We assume that $\rating = (slider - 50)/50 \in [-1,1]$, where $slider \in [0,100]$ is the position of the slider discussed in Section~\ref{sec:ratings}.
The variables $\localscore{\contributor \video}$ correspond to the (learned) score of video $\video$ given by contributor $\contributor$, while $\globalscore{\video}$ will be the global score of video $\video$.
Finally, for clarity, we denote $\datafamily$ and $\localscorefamily$ the tuple of data $\data{\contributor}$ and $\localscore{\contributor}$.

Tournesol then considers the following loss function:
\begin{equation}
    \Loss (\globalscore{}, \localscorefamily, \datafamily)
    \triangleq \sum_{\contributor \in [\Contributor]} \sum_{(\video, \videobis, \rating) \in \data{\contributor}} \lossperinput (\localscore{\contributor \video} - \localscore{\contributor \videobis}, \rating)
    + \collabweight \sum_{\contributor \in [\Contributor]} \sum_{\video \in [\Video]} \weight{\contributor \video} \absv{\localscore{\contributor \video} - \globalscore{\video}}
    + \globalweight \collabweight \sum_{\video \in [\Video]} \globalscore{\video}^2,
\end{equation}
where $\lossperinput$ is a loss-per-input function which will be detailed in the next section, $\collabweight$ is the weight of collaboration, $\weight{\contributor \video}$ is the weight of contributor $\contributor$ on the score of video $\video$, and $\globalweight$ is the weight of the regularization of the global scores.
A precise analysis of this loss function is beyond the scope of this paper, and is currently being investigated.

\begin{figure}[!ht]
    \centering
    \includegraphics[width=\linewidth]{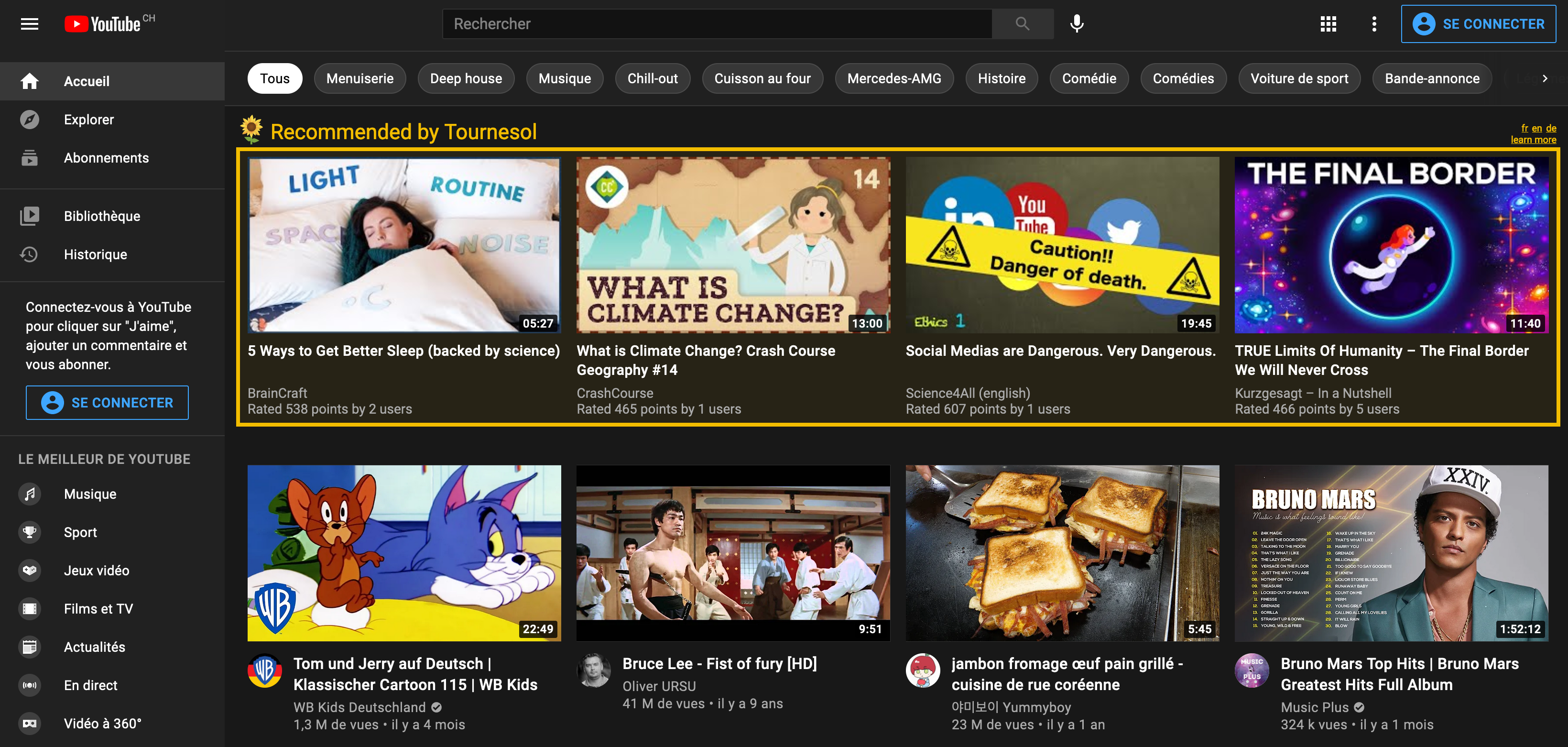}
    \caption{Using our Chrome or Firefox extensions, Tournesol users are provided with Tournesol recommendations, based on the global scores $\globalscore{}^*$.}
    \label{fig:recommendations}
\end{figure}

\begin{figure}[!ht]
    \centering
    \includegraphics[width=\linewidth]{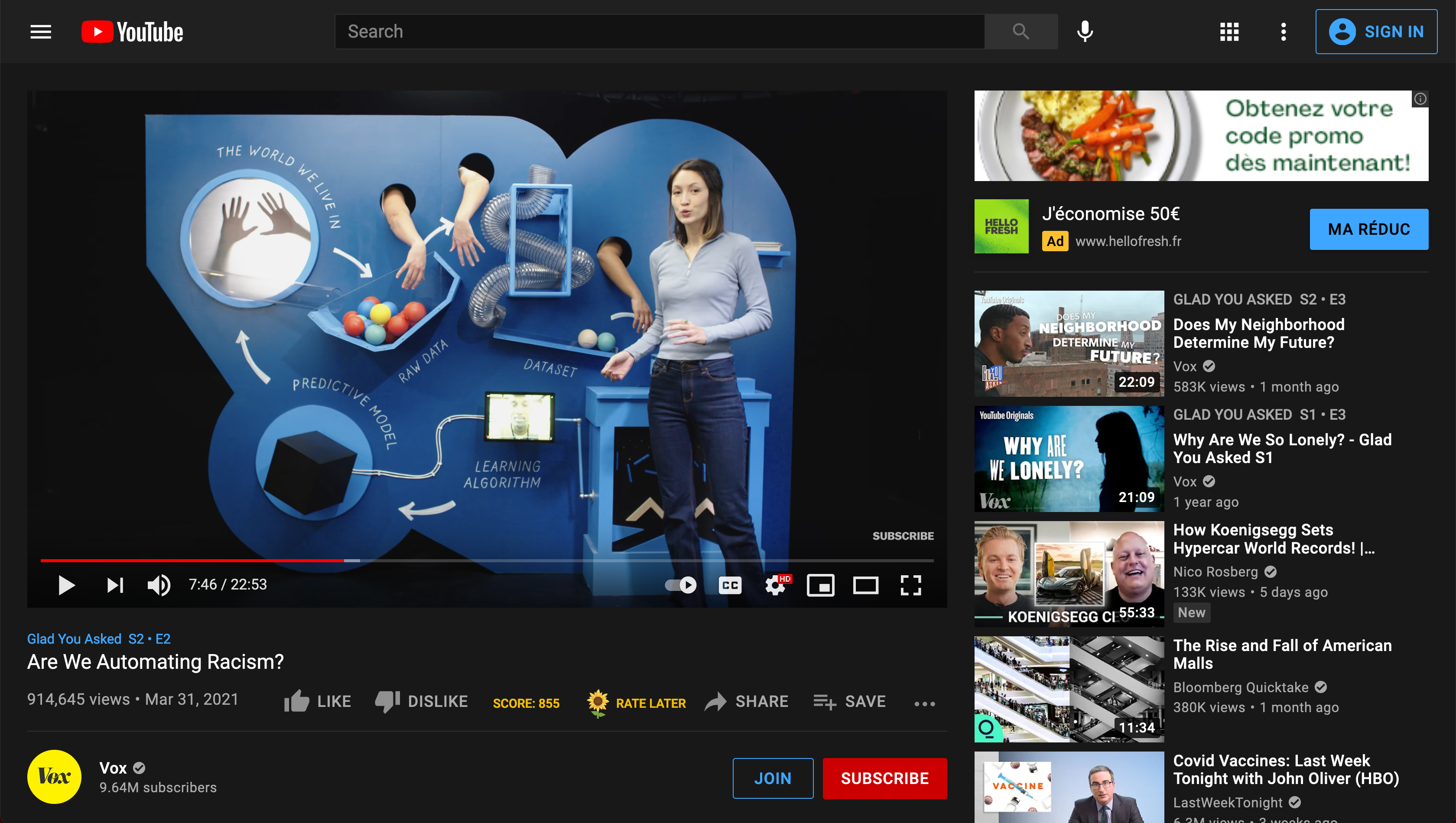}
    \caption{Using our Chrome or Firefox extensions, Tournesol users are provided with Tournesol's global scores $\globalscore{}^*$ on the video page, next to the like/dislike statistics. Contributors can also click on ``rate later'' to add the video to their Tournesol rate-later list.}
    \label{fig:score_on_video_page}
\end{figure}

Tournesol's individual scores $\localscore{\contributor \video}$ and global scores $\globalscore{\video}$ are then computed by minimizing the Tournesol loss function.
In particular, the global scores are used to make recommendations (see Figure~\ref{fig:recommendations}) and are displayed on Youtube, for users using on browser extension (see Figure~\ref{fig:score_on_video_page}).

\subsection{Bradley-Terry model}
\label{sec:bbt}

Let us now detail the loss-per-input function $\lossperinput$ currently used by Tournesol. 
We derive this loss from the classical Bradley-Terry model~\cite{BradleyTerry52}, which assumes that video comparisons are binary: either the video $\video$ is better than $\videobis$ (expressed by a rating $\rating{} = 1$), or vice versa ($\rating{} = -1$).
The probability of $\rating{} = 1$ (i.e., preferring video $\video$ to video $\videobis$) depends on the difference $t$ between the videos' scores: $t \triangleq \localscore{\contributor \video} - \localscore{\contributor \videobis}$.
The larger the difference $t$ is, the more probable it is that the contributor declares preferring $\video$ to $\videobis$.
The loss is then the negative likelihood of the data, assuming independence of different ratings, i.e.
\begin{equation}
    \lossperinput ( t, \rating{} ) \triangleq \ln \left( 1 + \exp(t \rating{}) \right).
\end{equation}
Note that this loss is also essentially the logistic regression loss.

The usual interpretation of this loss is that $\video$ will be one point above $\videobis$ (i.e., $t=1$), if it is $e$ times more likely to be rated better.
However, this model is no longer adequate when contributors provide continuous ratings $\rating{} \in (-1,1)$.
In fact, with the Bradley-Terry model, providing a rating $\rating{} = 0$ is equivalent to not contributing at all.
However, a rating $\rating{} = 0$ may be more accurately interpreted as the contributor  saying that the two videos should have similar scores.
The Bradley-Terry model fails to capture the nuanced information contained within continuous ratings.

This remark calls for future research to investigate alternative learning algorithms, which may yield better interpretability, and may better capture the intuitive idea that, when a contributor judges $\rating{} \approx 0$, they are actually saying that the two compared videos should have similar scores.

\subsection{Hyperparameters}

We ran preliminary experiments to understand the impact of the hyperparameters of our learning model.
Intuitively, $\collabweight$ measures how much the global scores are used to adjust each contributor's local scores.
Large values of $\collabweight$ mean that the global scores are regarded as a very reliable prior.
Smaller values of $\collabweight$ imply that the contributor's scores are likely to quickly diverge from the global scores.
In fact, $1/\collabweight$ is the order of magnitude of the prior standard deviation on contributor $\contributor$'s score for video $\video$, given the common score $\globalscore{\video}$.
Intuitively, the more diversity there is in contributor's judgments, the smaller the value of $\collabweight$ should be.
Future research will address how to dynamically adjust $\collabweight$.

The hyperparameter $\globalweight$ weighs the prior regularization on the scores. 
More importantly, its value determines the Byzantine resilience of our learning model~\cite{FarhadkhaniGH21}.
Larger values of $\globalweight$ imply that a single contributor can hardly affect the global scores.
As the number of contributors on Tournesol grows, we plan to decrease the value of $\globalweight$.

\begin{figure}[!ht]
    \centering
    \includegraphics[width=\linewidth]{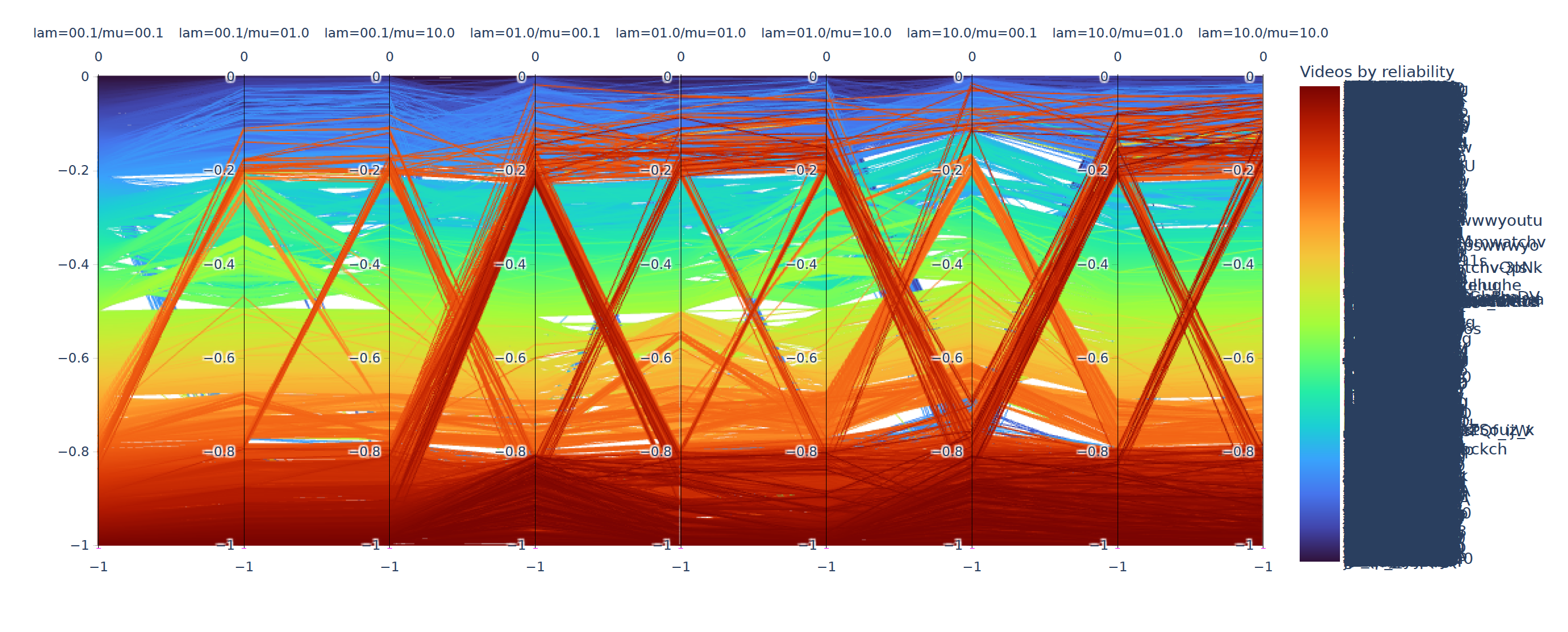}
    \caption{Video ranking of different videos based on the values of the hyperparameters $\collabweight$ and $\mu \triangleq \globalweight \collabweight$.}
    \label{fig:video_ranking}
\end{figure}

Figure~\ref{fig:video_ranking} shows how sensitive the scores of the videos are with respect to the hyperparameters.
This should not be surprising, given the heterogeneity of how often different videos have been rated.
In particular, a when $\globalweight = \mu / \globalweight$ is small, then the rating of a video cannot be large if the video was not rated by many contributors.
Further investigations of the impact of hyperparameters, and what a safe and ethical choice of hyperparameters should be, are needed.

Note however, that the effect of the hyperparameter $\globalweight$ vanishes in the limit of a large number of contributors. At this point, the global scores are simply the medians of the contributors' scores. 
Likewise, the effect of the hyperparameter $\collabweight$ also vanishes in the limit of a large number of data per contributor.
In other words, if Tournesol is used by a large number of very active contributors, then the choice of the hyperparameters hardly matters.

Currently, the platform uses the values $\collabweight = \globalweight = 1$.
Finally, we defined $\weight{\contributor \video} \triangleq \Rating_{\contributor \video} / (\weightConstant + \Rating_{\contributor \video})$, where $\Rating_{\contributor \video}$ is the number of times contributor $\contributor$ rated video $\video$ against other videos. 
This allows to favor contributors who provided more ratings to a video.
We set the hyperparameter $\weightConstant = 3$.
As a result, a contributor would obtain half of their maximal influence by rating a video three times, and $75\%$ of it by rating it nine times.

\subsection{Non-verified contributors}

Note that, given that ratings are sparse, inferring individual scores from individual ratings only is suboptimal. In particular, this approach is vulnerable to Stein's paradox~\cite{stein1956,JamesStein1961}.
Instead, to learn individual scores, even for non-verified contributors, we leverage the learned global scores $\globalscore{}^*$ (note that minimizing $\Loss$ does this automatically for verified contributors).
This corresponds to minimizing this loss for every non-verified contributor $\contributor$:
\begin{equation}
    \Loss_\contributor (\localscore{\contributor}, \data{\contributor}) 
    \triangleq \sum_{(\video, \videobis, \rating) \in \data{\contributor}} \lossperinput(\localscore{\contributor \video} - \localscore{\contributor \videobis}, \rating) 
    + \collabweight \sum_{\video \in [\Video]} \weight{\contributor \video} \absv{\localscore{\contributor \video} - \globalscore{\video}}.
\end{equation}
Such learned scores are used to provide individual recommendations on the non-verified contributors' Tournesol page.

\subsection{Customizable recommendations}

Tournesol applies the learning algorithm described above to the different quality criteria. 
This allows users to obtain customizable recommendations and search results, by adjusting the importance they assign to the different quality criteria.
The videos are then ranked based on the weighted sum of the scores per criteria, where the weights are given by the positions of the user's sliders (see Figure~\ref{fig:top_recommendations}).
However, we leave open the question of designing personalized robustly beneficial content recommendation, as discussed in Section~\ref{sec:alignment}.

\subsection{Strategyproofness}

We stress that the Licchavi framework with such $\ell_1$ collaborative regularizations has been tied with strategyproofness theorems in~\cite{FarhadkhaniGH21}.
We acknowledge, however, that the strategyproofness theorem in~\cite{FarhadkhaniGH21} does not quite apply to our setting though, partly because our loss-per-inputs are not actually gradient-PAC, but more importantly because our queries are not coordinate-wise.
Nevertheless, our framework applies the fundamental fairness principle ``one voter, one unit force'' proposed by~\cite{ElMhamdiFGH21}, which suggest a reasonable amount of strategyproofness (and of Byzantine resilience).
Future research is however needed to better determine the extent to which our algorithms are robust and strategyproof.

\section{Statistics}
\label{sec:statistics}

To highlight the value of our database, we present insightful preliminary data analyses.
Evidently, our data will evolve, hopefully drastically, in the coming months.

\subsection{Contributors' contributions}

\begin{figure}[!ht]
    \centering
    \includegraphics[width=.8\linewidth]{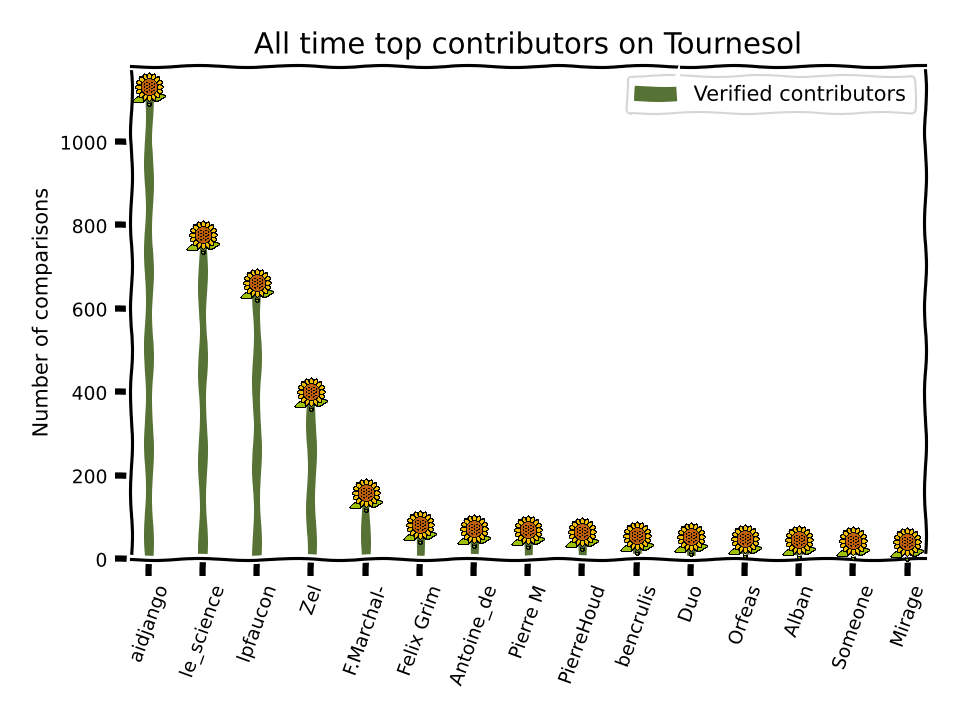}
    \caption{This figure displays the top contributors on Tournesol, and their number of ratings.}
    \label{fig:contributions}
\end{figure}

Figure~\ref{fig:contributions} displays the number of contributions per user among the top current contributors.
Perhaps unsurprisingly, this statistics seems to follow a heavy tail distribution, with a few contributors providing most of the ratings, and the majority of them providing very few ratings.
Fortunately, our learning algorithm, based on Licchavi with norm-based collaborative regularizations, fits the principle ``one voter, one unit force''~\cite{ElMhamdiFGH21}, and thus prevents the leading contributors from having an uncontrollable influence on global scores.

\subsection{Connectivity of the data}

\begin{figure}[!ht]
\centering
    \begin{subfigure}{.49\textwidth}
        \centering
        \includegraphics[width=\linewidth]{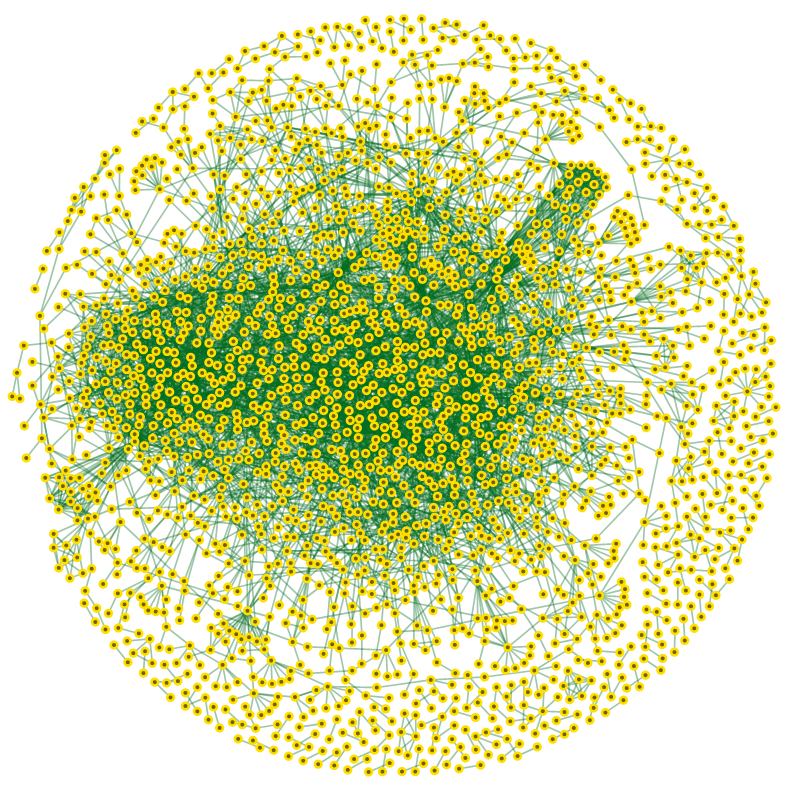}
    \end{subfigure}
    \begin{subfigure}{.49\textwidth}
        \centering
        \includegraphics[width=\linewidth]{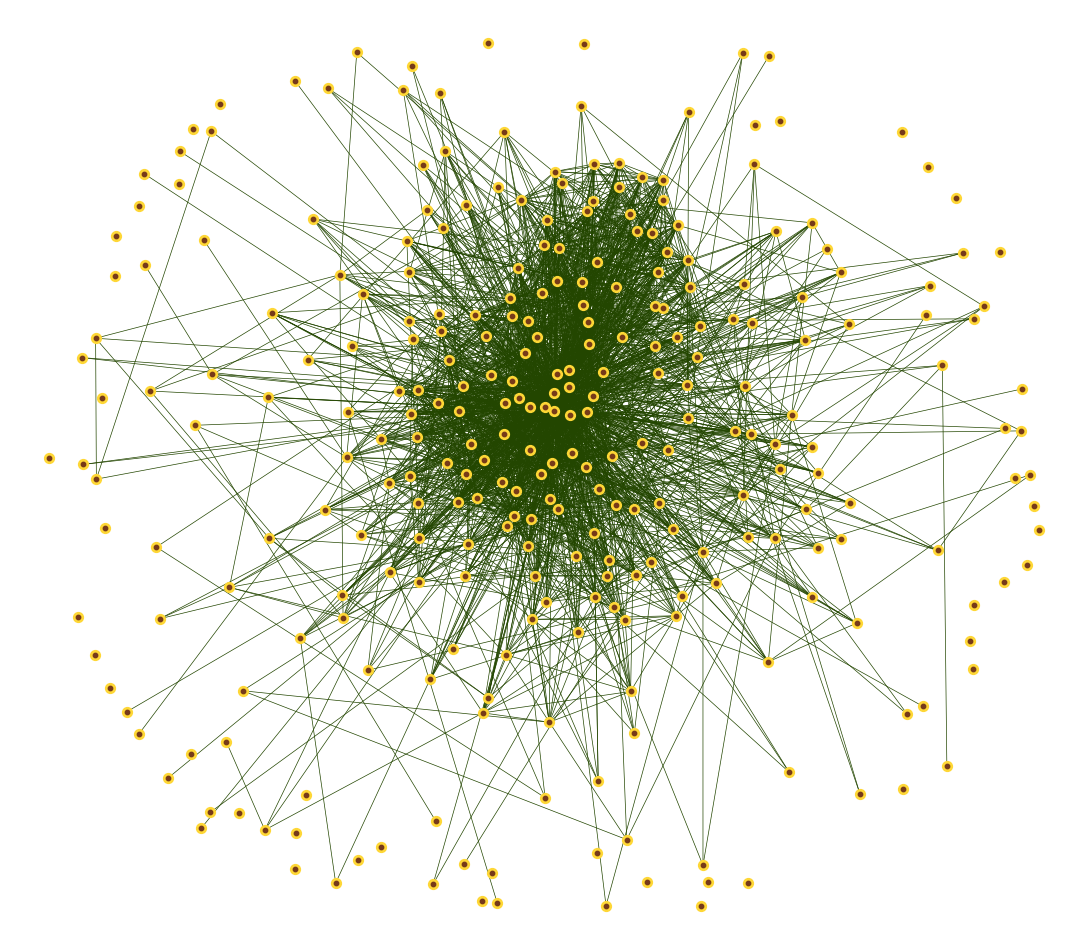}  
    \end{subfigure}
    \caption{On the left, each sunflower on the figure represents a video, and two videos are connected if there is a contributor who compared them at least once. On the right, each sunflower on the figure represents a contributor, and two contributors are connected if there is a video that they both rated.}
    \label{fig:comparison_graph}
\end{figure}

The left part of Figure~\ref{fig:comparison_graph} shows the connectivity of the graph of video comparisons.
In this graph, two videos are connected if they have been compared at least once on the Tournesol platform.
While the graph features a dense center, many videos are only weakly connected to most other videos.
The connectivity of the graph is important, as it allows to guarantee that all videos are scored along the same scale.

The right part of Figure~\ref{fig:comparison_graph} displays the commonalities between contributors. 
Namely, nodes on the graph are contributors, and two contributors are connected if their lists of rated videos have a non-empty intersection.

\subsection{Correlations between criteria}

\begin{figure}[!ht]
\begin{subfigure}{.49\textwidth}
  \centering
  \includegraphics[width=.95\linewidth]{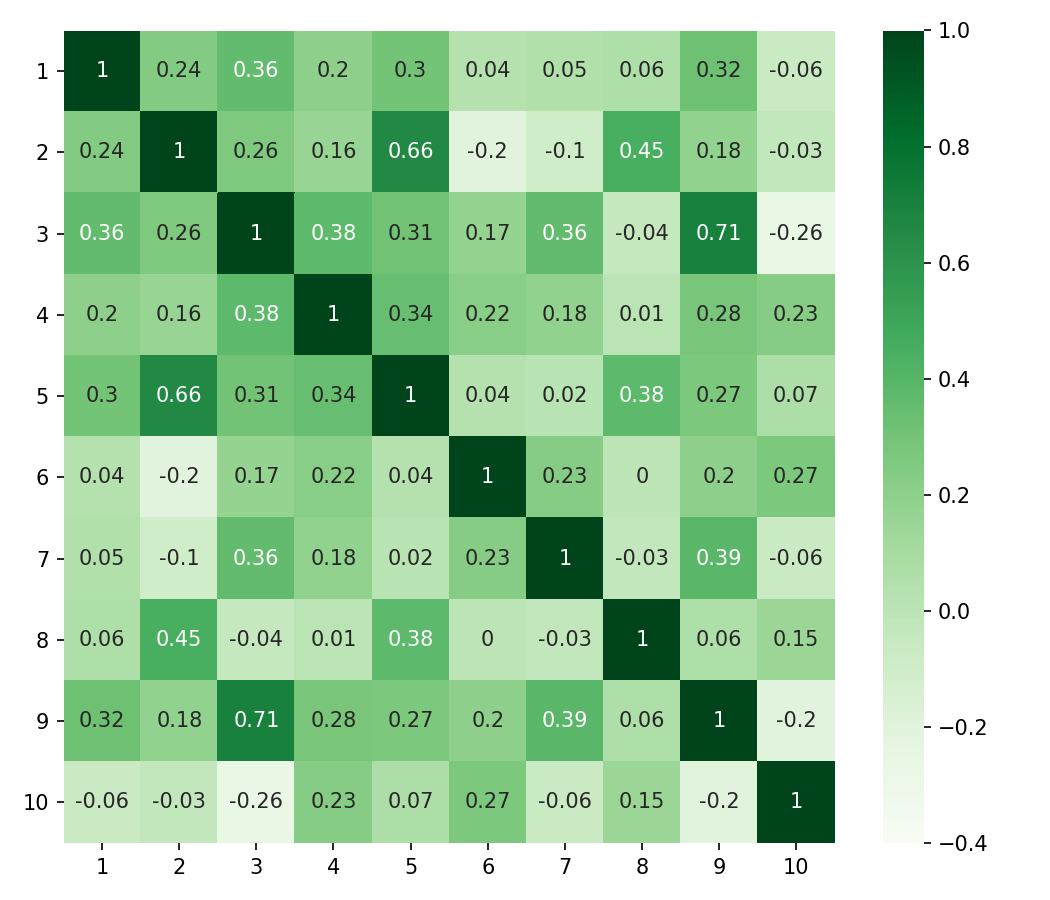}  
  \caption{All videos rated on Tournesol}
  \label{fig:scores_correl_all}
\end{subfigure}
\begin{subfigure}{.49\textwidth}
  \centering
  \includegraphics[width=.95\linewidth]{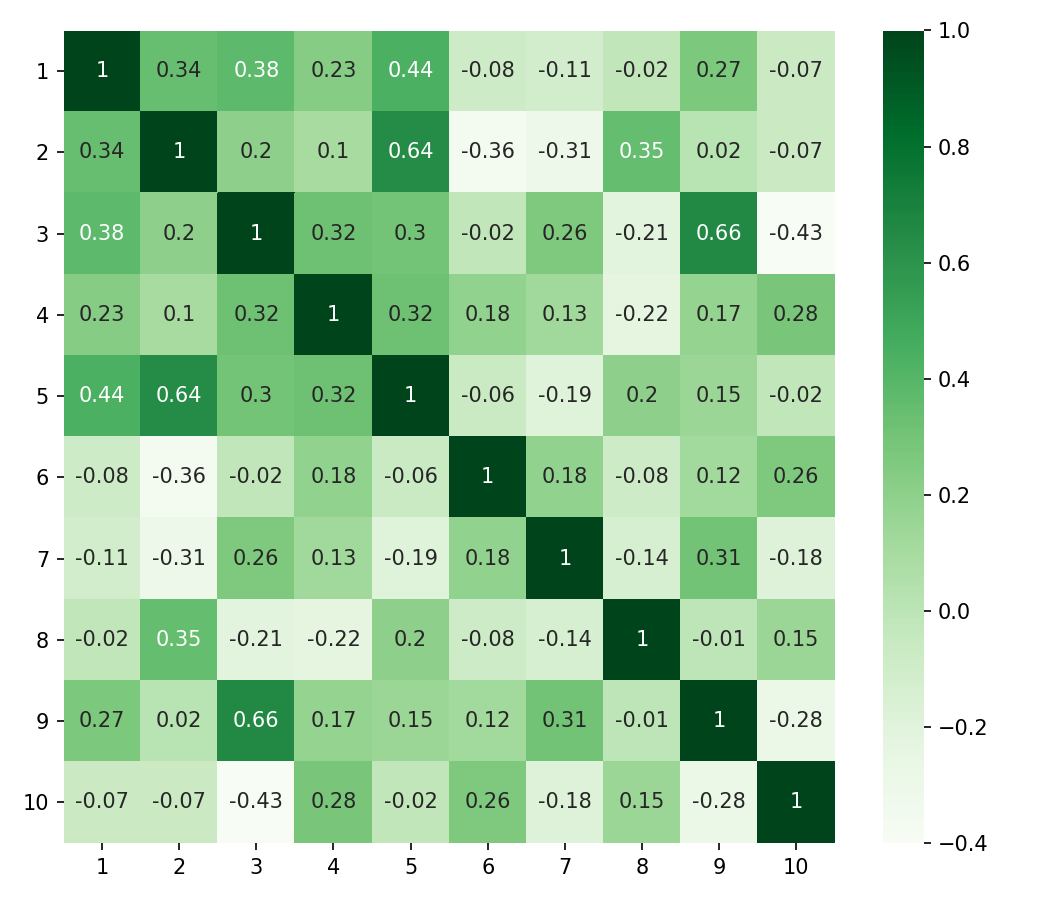}  
  \caption{Top 10\% videos rated on Tournesol}
  \label{fig:scores_correl_top10percent}
\end{subfigure}
\caption{Correlations between quality criteria, whose numbering are given in Table~\ref{table:criteria}. Ideally, a set of criteria should be orthogonal enough to provide complementary information with minimal effort on the contributor's side.}
\label{fig:scores_correl}
\end{figure}

Figure~\ref{fig:scores_correl_all} reports the correlations between quality criteria. 
It shows that the choice of our criteria is somewhat reasonable, as most criteria are only weakly correlated.
Some criteria are however somewhat redundant like ``reliable and not misleading'', ``clear and pedagogical'' and ``resilient to backfiring risks''.
Similarly, ``important and actionable'' and ``encourage better habits'' are also quite correlated.

As expected given Berkson's paradox~\cite{Berkson46}, the correlations decrease if we only consider the top 10\% videos on Tournesol (Figure~\ref{fig:scores_correl_top10percent}).
This is important as these are the videos that Tournesol users will be more exposed to.

\begin{table}[!ht]
\centering
\small
\begin{tabular}{@{}ll@{}}
\toprule
\# & Name of criteria               \\ \midrule
1  & Should be largely recommended  \\
2  & Reliable and not misleading    \\
3  & Important and actionable       \\
4  & Engaging and thought-provoking \\
5  & Clear and pedagogical          \\
6  & Layman-friendly                \\
7  & Diversity and Inclusion        \\
8  & Resilience to backfiring risks \\
9  & Encourages better habits       \\
10 & Entertaining and relaxing      \\ \bottomrule
\end{tabular}
\caption{Numbering of the criteria used in Figure~\ref{fig:scores_correl}.}
\label{table:criteria}
\end{table}

\subsection{Rating patterns}
\label{sec:rating_patterns}

\begin{figure}[ht]
\begin{subfigure}{.5\textwidth}
  \centering
  \includegraphics[width=.9\linewidth]{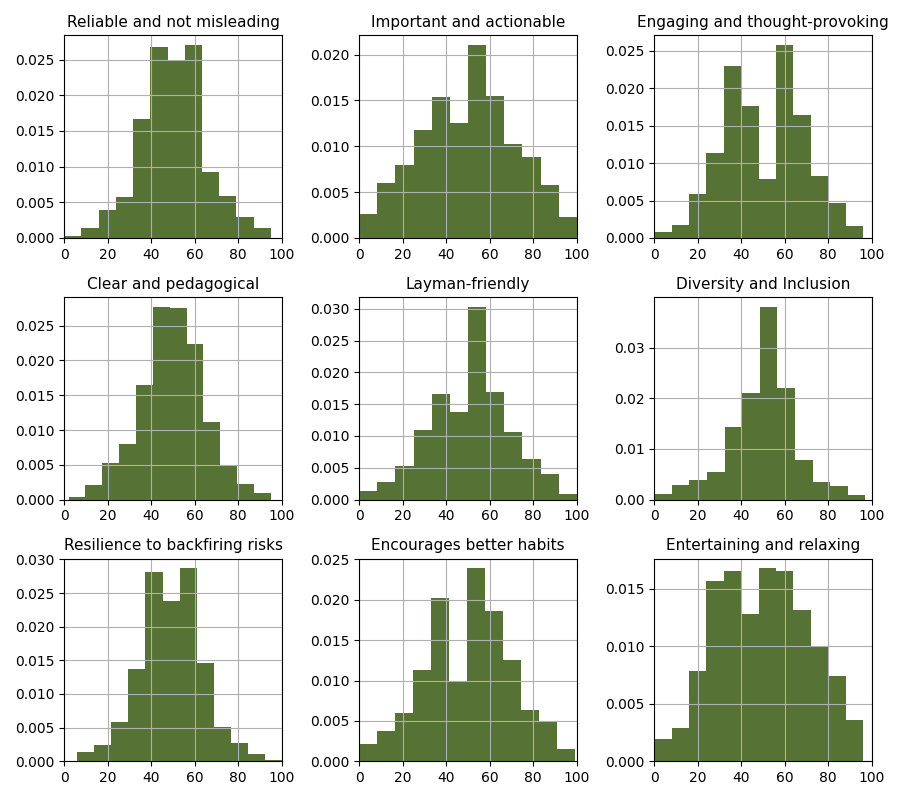}  
  \caption{Contributor ``aidjango''}
  \label{fig:cursor_histogram_a}
\end{subfigure}
\begin{subfigure}{.5\textwidth}
  \centering
  \includegraphics[width=.9\linewidth]{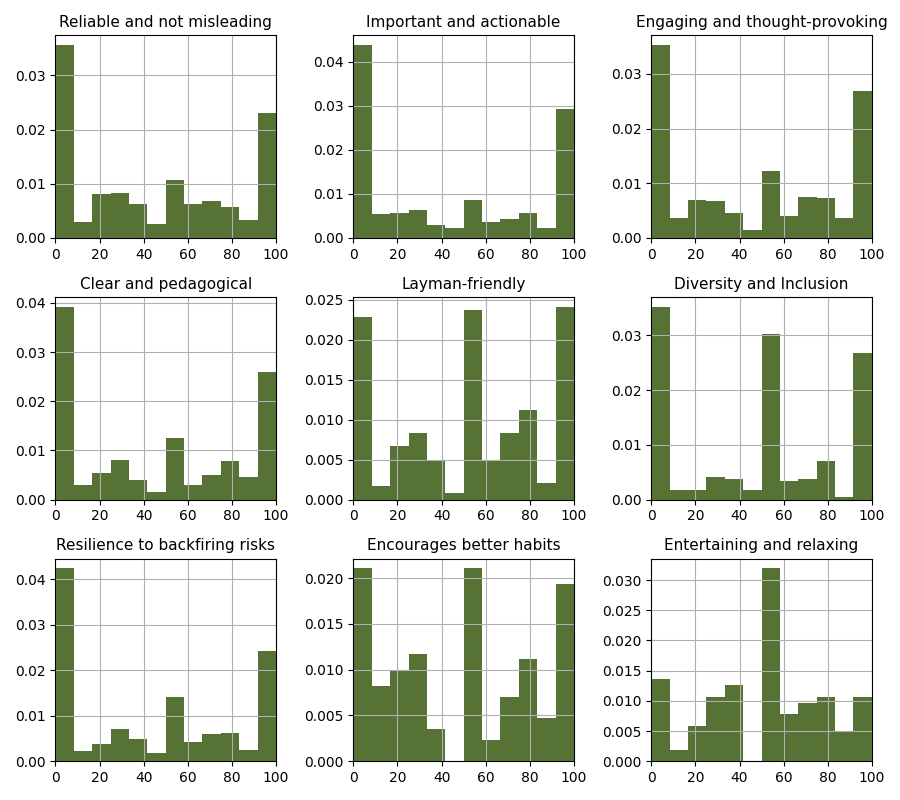}  
  \caption{Contributor ``le\_science4all''}
  \label{fig:cursor_histogram_b}
\end{subfigure}
\caption{Distribution of the ratings for each criteria, showing difference of rating patterns between two Tournesol contributors.}
\label{fig:cursor_histogram}
\end{figure}

As it is not formally defined how contributors should rate a pair of videos, we expected many different rating styles. 
Figure~\ref{fig:cursor_histogram} shows how two contributors whose rating patterns are drastically different, despite rating similar videos. 
Namely, while contributor ``aidjango'' provided ratings close to ``indifferent'', contributor ``le\_science4all'' provided more extreme ratings.
This suggests that the discrepancies between their individual scores will be due to their rating style, rather than actual differences in their judgments.
As discussed in Section~\ref{sec:correcting_rating_patterns}, research should address how to adapt the learning model to contributors' different rating patterns.

\subsection{Pareto front}

\begin{figure}[!ht]
    \centering
    \includegraphics[width=.7\linewidth]{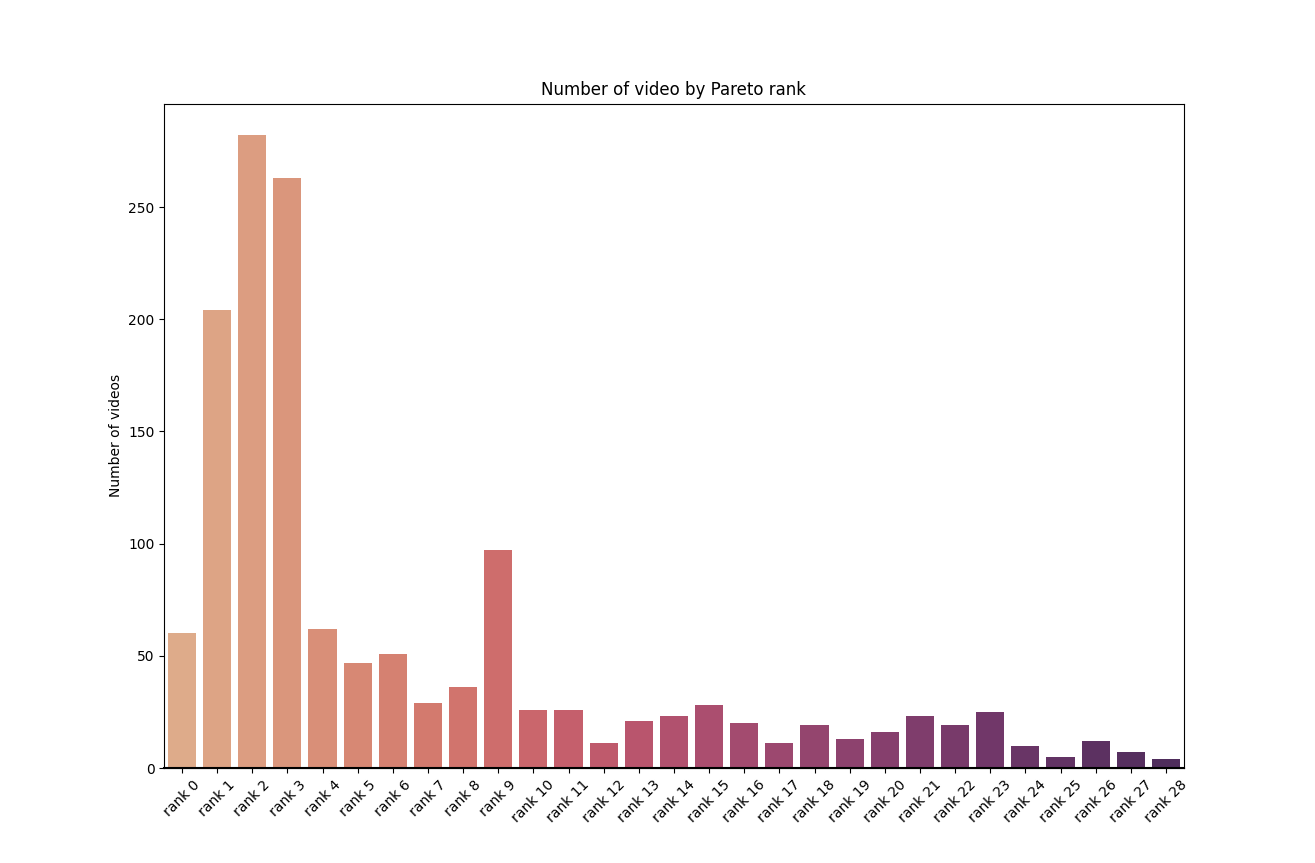}
    \caption{Number of videos per Pareto rank.}
    \label{fig:pareto_rank}
\end{figure}

Figure \ref{fig:pareto_rank} shows the number of videos per Pareto rank. To recall, the Pareto rank of a video is the number of videos that need to be removed so that the video becomes a Pareto-optimal video, i.e., a video that no other video outperforms on all quality criteria.

We note that, currently, there is a large number of pareto-optimal videos ($>50$) because there are 10 dimensions allowing for plenty of trade-offs. 
This highlights the usefulness of customizable recommendations.
It is also noteworthy that many videos have rank 3 or less. This is likely due to the fact that current contributors mostly rate videos that they deem particularly worthy of recommendation. 
It will be interesting to see how these statistics evolve as our database grows, and contains a more diverse range of video rating styles.

\section{Research challenges}
\label{sec:challenges}

Tournesol raises numerous fascinating research challenges. Below, we sketch some of these challenges. 
We would welcome any potential collaborations on any of the following challenges, and actively encourage the use of our public database for good, and, in particular, for research purposes.

\subsection{Aggregate the different criteria into a score}

We expect the combination of many different quality criteria to yield a more reliable judgment of what content ought to be recommended at scale, or to a given a specific user.
However, the appropriate aggregation of our different quality criteria is still unclear, especially given probable nonlinear phenomena.
For instance, a reasonably reliable content seems vastly superior to a misleading content, whereas an extremely reliable content does not seem vastly superior to a reasonably reliable content.
We hope to investigate how best to do this, by comparing aggregations of optional quality criteria to the ``Should be largely recommended'' criterion, and by factoring in discrepancies between different contributors' ratings, given their expertise and contributing profiles.

\subsection{Verifying more contributors}
\label{sec:pop}

We hope to leverage vouching to verify more contributors in a Byzantine-resilient manner.
However, it is unclear so far what are the best algorithms to achieve this.
We are currently investigating this relatively general problem (see Section~\ref{sec:vouch}), which is tightly connected to ideas such as \emph{Web of Trust}~\cite{Mueller21} or reputation mechanisms~\cite{OliveiraRMCOM20}.

\subsection{Debiaising the contributing population}

Unfortunately, like in many online participatory projects~\cite{Awad18}, we expect huge participation imbalances.
Young male technology-savvy individuals are probably more likely to participate in Tournesol, which means that they will be overrepresented in the Tournesol database.
We are currently investigating how to leverage our demographic data to debias the Tournesol recommendations, e.g., by giving stronger voting rights to individuals whose communities are underrepresented in the Tournesol database.

\subsection{Learning and correcting rating patterns}
\label{sec:correcting_rating_patterns}

The current ``binomial Bradley-Terry'' model assumes that all contributors rate in the same manner, given implicit score differences.
In practice, as discussed in Section~\ref{sec:rating_patterns} and as depicted by Figure~\ref{fig:cursor_histogram}, different contributors will have very different rating habits.
To make sure that the learned individual scores are fairly comparable, the rating model should be hyperparameterized, and the hyperparameters should be learned and adjusted per contributor.
We are currently investigating ways to do this.

\subsection{Leveraging expertise}

On technical topics like vaccination or climate change, especially when misconceptions are widespread in the general population, it seems desirable to assign more voting rights to experts, especially when judging the reliability of content within their domains of expertise.
This issue is intimately connected to Condorcet's jury problem~\cite{condorcet1785,nitzan1982}.
We are currently investigating how best to leverage personal information to determine appropriate voting rights.

\subsection{Understanding and improving contributors' experiences}

To understand the reliability of Tournesol's data, it is critical to understand the psychology of contributors when they provide judgments.
We are currently investigating contributors' thought processes, and how rating on Tournesol affects contributors' reasonings.
This work is inspired by~\cite{LeeKKKYCSNLPP19}, who interviewed participants of their participatory ethical algorithm design process, and found out that, by the participants' own assessments, this enabled them to improve their ethical judgments and to increase their trust in the participatory system.

\subsection{Active learning}

To increase the quality and the quantity of the database, it is crucial for Tournesol to help contributors select the videos to compare in the best possible ways.
This is a challenging task, as it encompasses several considerations.
On one hand, the comparison should provide valuable data, by typically querying interesting comparisons, involving videos with too few ratings or better connecting the graph of pairwise video comparisons.
On the other hand, it is important that this demands as little cognitive investment from the contributor as possible. 
This typically requires suggesting the contributor to compare videos that they have recently watched and assessed.
Finding out how to best do this is a research challenge that we hope to investigate soon.

\subsection{Volition learning}
\label{sec:volition}

Despite our efforts, we cannot expect the Tournesol database to contain fully reliable human judgments. 
We expect many of ratings will be provided by contributors who might not be considering all the possible ramifications and unwanted side effects of promoting a video content at scale when they provided their judgments.
In particular, some judgments will arguably be more reliable than others.
Understanding which judgments are more reliable than others is a vast research program that is critical to design safer and more ethical algorithms.
More reliable judgments are sometimes called \emph{volitions}, rather than \emph{preferences}.
Volitions are also often described as second-order preferences: they correspond to what we would prefer to prefer, rather than what we instinctively prefer~\cite{tarleton10,HoangElMhamdi19}.
We hope to initiate the research on \emph{volition learning}, by leveraging the meta-data of our database (see Section~\ref{sec:meta-data}).

\subsection{Privacy-preserving learning}

While we believe that they have a very reasonable amount of privacy protection, our current algorithms require a deeper analysis to understand the extent to which they guarantee the protection of private ratings.
Future research should also investigate how to strengthen privacy without harming too much the quality and the security of the Tournesol scores, and how to leverage private personal information about our contributors (which are currently not used) in a privacy-preserving manner.
Perhaps most importantly, ideally, Tournesol would be able to leverage private ratings to score videos without being a single point of failure for private data protection.
Moving forwards, a challenging research avenue is to investigate the extent to which the mission of Tournesol can be achieved without any transfer of private information.

\subsection{Decentralize Tournesol}

A longer-term goal is to fully decentralize Tournesol. 
In this vision, the data would no longer be stored on Tournesol's server, but would be replicated appropriately on a large number of contributors' devices.
Moreover, the computations of Tournesol scores should also be decentralized, while guaranteeing Byzantine resilience.
Recent research in fully decentralized Byzantine learning has provided the building blocks of such a decentralization~\cite{collaborative_learning}, but more research is needed to understand how to best do so in the context of Tournesol.

\subsection{Generalizing learning}

Right now, Tournesol only leverages the Tournesol database to compute individual and global scores.
However, in the longer-term, it seems desirable to leverage additional information, such as the videos' channels or their descriptions, to generalize the individual scores of a contributor to videos that they have not rated (and not even watched!).
In particular, the use of natural language processing on video captions can be a promising avenue, though this probably requires a larger database than the current database.

\subsection{Enable multiple individual learning algorithms}

Right now, Tournesol is imposing each contributor to use our model, and in particular the binomial Bradley-Terry loss (see Section~\ref{sec:bbt}).
However, especially as we move to algorithms able to generalize to videos the contributor has not watched yet, it seems desirable to enable the contributor to choose which learning model they want to use.
To achieve this, Licchavi must be adapted to enable the collaboration of different local learning models.

\subsection{Ethical language models}

In the long run, we hope that Tournesol's database will be useful to design more sophisticated ethical algorithmic products, such as a ethical language models~\cite{BenderGMS21}.
Typically, when prompted, such language models should consistently produce texts that are robustly beneficial to communicate, either at scale, or to targeted consumers.
Determining how to combine large language models~\cite{FedusZS21} with Tournesol's database to design safe and ethical language models is currently a very open research challenge.

\subsection{Leveraging Tournesol's scores for alignment}
\label{sec:alignment}

For most large-scale algorithmic systems, such as recommendation algorithms, the objective function is the main avenue we have to make them robustly beneficial. 
The problem of making this objective function safe and desirable to optimize is called the \emph{alignment problem}~\cite{yudkowsky16,russell2019human,hoang19}.
While a large, secured and trustworthy database of reliable human judgments seems critical to solve alignment, it is unclear so far how to best leverage the Tournesol database to achieve this, especially while being personalized~\cite{FarhadkhaniGH21} and while being robust to Goodhart's law~\cite{ElMhamdiHoang21}.
This is arguably the most challenging and the most exciting of all the problems Tournesol raises.

\section{Conclusion}
\label{sec:conclusion}

\paragraph{Summary.} In this paper, we introduced Tournesol, a platform whose goal is to collect and curate a large, secured and trustworthy database of reliable human judgments.
We discussed the main features of the platform, the algorithm we use today to leverage this database to make more robustly beneficial recommendations, and the numerous research challenges that must be overcome to make Tournesol a success.
We strongly believe that the creation of such a database will stimulate research and development on ethical algorithms, which should help improve the informational diet of billions of people for the better. 
To achieve this ambitious goal, however, we will need support. 

\paragraph{Call for action.} To date, many individuals have voluntarily contributed to our code base and our database, for which we are immensely grateful. However, we still need a large amount of software development to make our platform more contributor-friendly, more secure and more scalable. For this reason, the Tournesol Association is currently searching for funding to hire top developers to maintain and develop the platform. 
Just as critically, Tournesol needs to attract a large number of contributors and to obtain human judgments from a diverse population. 
We would also be hugely grateful for the support of fellow researchers and institutions. 
To help us, \href{https://tournesol.app/}{try out the platform} and please spread the word! Your contributions will be improving the quality of online discourse for everyone.

\bibliographystyle{alpha}
\bibliography{references}

\end{document}